\documentclass[twocolumn]{aastex7}

\begin{document}

\title{Spectro-Polarimetric Study of Weakly Magnetized Neutron Star X-ray binary GX 349+2}

\correspondingauthor{Raj Kumar}
\author[0000-0002-0786-7307]{Raj Kumar}
\affiliation{Astrophysical Sciences Division, Bhabha Atomic Research Centre, Mumbai - 400085, India}
\affiliation{Homi Bhabha National Institute, Mumbai - 400094, India}
\email[show]{arya95raj@gmail.com}  

\author[0000-0001-6877-4393]{SUBHASISH DAS}
\affiliation{Department of Pure and Applied Physics, Guru Ghasidas Vishwavidyalaya (A Central University)\\ Bilaspur (C. G.)- 495009, India}
\email[show]{subhasishboson@gmail.com }


\begin{abstract}

We report the first spectro-polarimetric investigation of the bright Z-type source GX 349+2 using simultaneous observations of \textit{IXPE}, \textit{NuSTAR} and \textit{Swift/XRT}. The source exhibited significant polarization in the 2-8 keV energy range during the flaring branch (FB) and normal branch (NB). The estimated polarization degree (PD) and polarization angle (PA) for FB are $1.74 \pm 0.52\%$ ($3.3\sigma$) and $19.4 \pm 8.9^\circ$, respectively; while for NB, PD and PA are $0.8 \pm 0.22\%$ ($3.6\sigma$) and $35.4 \pm 7.9^\circ$, respectively. 
The energy-resolved polarization for NB revealed an increase in PD from $0.78\pm0.2\%$ ($3.6\sigma$)  to $1.32\pm0.40\%$ ($3.3\sigma$) and a change in PA from $17.9\pm8.1^\circ$, and $53.2\pm8.6^\circ$, in the energy range of 2-4 and 4-8 keV, respectively.
Using the simultaneous observations of \textit{Swift/XRT} and \textit{NuSTAR}, we investigated the spectral properties of the source during NB, and the Western model well explained it. The spectra also depicted a strong and broad Fe K$\alpha$ line. 
However, spectro-polarimetric analyses carried out by \textit{IXPE} align closely with model-independent polarimetric results.
We discuss results obtained from the polarimetric studies in the context of various coronal geometries, and we confirm the slab-like geometry in the NB for GX 349+2.
\end{abstract}



\section{Introduction} \label{sec:intro}
The weakly magnetized neutron star low-mass X-ray binaries (NS-LMXBs) are classified as atoll-type and Z-type sources \citep{Hasinger1989}. The Z-type sources trace the Z shape on the color-color diagram (CCD) in the time scale of one day to weeks, while atoll sources trace a C-type pattern on a time scale of weeks to months on the CCD. The Z-track is primarily divided into three main branches, namely, the horizontal branch (HB), the normal branch (NB), and the flaring branch (FB). The transition between HB and NB is known as hard apex (HA), while the transition between NB and FB is known as soft apex (SA). Z sources are classified as either Cyg-like or Sco-like. Cyg-like ones have strong HB and NB but a weak FB, while Sco-like ones exhibit strong NB and FB with a faint HB.

The X-ray emission of NS-LMXBs is often described as a combination of a soft component and a hard component. The soft component originates due to the thermal emission from the NS surface (either from the boundary layer or spreading layer ) or a multicolor disc. However, the hard component is caused by inverse Compton scattering of soft photons by the hot electron cloud/corona. 
The spectra are commonly described by two primary models: the ``Eastern model" \citep{Mitsuda1984, Mitsuda1989} and the ``Western model" \citep{White1988}. 
In the Eastern model, the soft component is produced by a multicolor accretion disk, whereas the hard component emerges from the Comptonization of thermal emission from the NS. 
In contrast, the Western model considers the soft component to be caused by a single temperature blackbody from the NS surface, whereas the hard component is generated by the Comptonization of photons supplied from the accretion disc.
The above-mentioned models were proposed a few decades ago, but due to their strong spectroscopic degeneracy \citep{Barret2001}, the debate over coronal geometry persists till today. \cite{Cocchi2011} suggested that both seed photons from the multicolor disc and NS surface could undergo scattering in the corona. Moreover, the origin of the corona and its geometry remains an open challenge \citep{Degenaar2018}.

GX 349+2 is an SCO-like source that exhibits only NB and FB, which makes it unique from other Z-sources \citep{Hasinger1989, Kuulkers1998}. The estimated distance of GX 349+2 is most likely between 5 and 10 kpc \citep{Cooke1991, vanParadijs1994, Christian1997}. \citep{Iaria2009} estimated the inclination angle of GX 349+2 to be between $40^\circ$ and $47^\circ$ .
Many studies have been conducted throughout the years to better understand the emission process from GX 349+2. 
Several investigations favor the Eastern model \citep{Disalvo2001, Iaria2009}, while studies by \cite{Church2012} and \cite{Ding2016} support the Western model.

Previous studies revealed the presence of a relativistic Fe-K$\alpha$ emission line in the spectrum of this source \citep{Cackett2008, Iaria2009, Coughenour2018, Kashyap2023}.

However, spectral analysis alone usually leads to ambiguity when deciding the best-fit models, emphasizing the need for additional constraining observables such as polarization to resolve such degeneracies.
With the launch of Imaging X-ray Polarimetry Explorer \textit{(IXPE)} \citep{Weisskopf2016, Weisskopf2022}, the polarimetric properties of different astrophysical objects in X-ray have been extensively researched. Polarimetric studies of astrophysical objects have increased our understanding of these systems.
Recent investigations have discovered concrete evidence of low PD in NB and FB, ranging from $0.6-2.0\%$ (Cyg X-2: \cite{Ferinelli2023}, XTE J1701-462: \cite{Cocchi2023}, GX 5-1: \cite{Fabiani2024}, Sco X-1: \cite{LaMonaca2024}, Cir X-1: \cite{Rankin2024}, and GX 340+0: \cite{Bhargava2024}). \cite{Ursini2024} summarized the first polarimetric results from \textit{IXPE} on NS-LMXBs. 

Our main goals are to estimate the polarimetric properties of GX 349+2 and gain insight into the coronal geometry. This letter is organized as follows. Section \ref{sec:data} is dedicated to observations and data reduction of observations used in this work. Section \ref{sec:results} provides a concise analysis of GX 349+2 using \textit{IXPE}, \textit{Swift/XRT}, and \textit{NuSTAR}. We discussed and concluded our findings in section \ref{sec:dis}.

\section{Observation and data reduction} \label{sec:data}

\subsection{\textit{IXPE}}
\textit{IXPE} is an X-ray telescope dedicated to polarimetric imaging and spectral studies of celestial objects in the 2-8 keV energy range. It consists of 3 detector units. \textit{IXPE} observed the GX 349+2 from 2024 September 6 to 2024 September 8 (obsid: \texttt{03003601}) with an exposure time of 95 ks.
We used \texttt{HEASoft-V6.33.2}, different tools (like \texttt{xpbin} and \texttt{xpeselect}) of \texttt{ixpeobssim} software (v30.2.2), and calibration files corresponding to epoch 2024 January 25 (\texttt{obssim20240125 alpha075}) to analyze the data. To extract the source photons, we used a circular region of 100 arcsec. Due to the high count rate of the source, which exceeds 2 $cps/arcmin^2$, we do not apply any background rejection or subtraction \citep{Dimarco2023}. For model-independent polarimetric properties, we used the \texttt{PCUBE} task of the \texttt{xpbin} tool. The PHA spectra for \textit{I}, \textit{U}, and \textit{Q} Stokes parameters were extracted from \textit{IXPE} level 2 event lists using the \texttt{XPBIN} task. We used a constant energy binning of 120 eV for all three Stokes parameter energy distributions (\textit{I}, \textit{U}, and \textit{Q}).

\subsection{\textit{Swift/XRT}}
The \textit{Swift/XRT} \citep{Burrows2005} observed the GX 349+2 on 2024 September 08 (obsid: \texttt{00089970001}) in WT mode with an exposure time of 1.9 ks \citep{Gehrels2004}. The \texttt{xrtpipeline} was used to reprocess and filter the data. The count rate in the circular region with 30-pixel radii centered at the source position is greater than 100 cts/s, indicating data is piled up \citep{Romano2006}. So, we used an annular region with 30-pixel and 3-pixel outer and inner radii, respectively. The \texttt{XSELECT} task was deployed to extract the spectrum and lightcurve files. Since the source is bright, we have not extracted the background spectrum. The ARF was generated using the \texttt{xrtmkarf} task. The RMF file is obtained from CALDB files. Then we used the \texttt{grppha} task to add 3\% systematic error and group the minimum 30 counts per bin.\\

\begin{figure*} []
    \centering
    \gridline{\fig{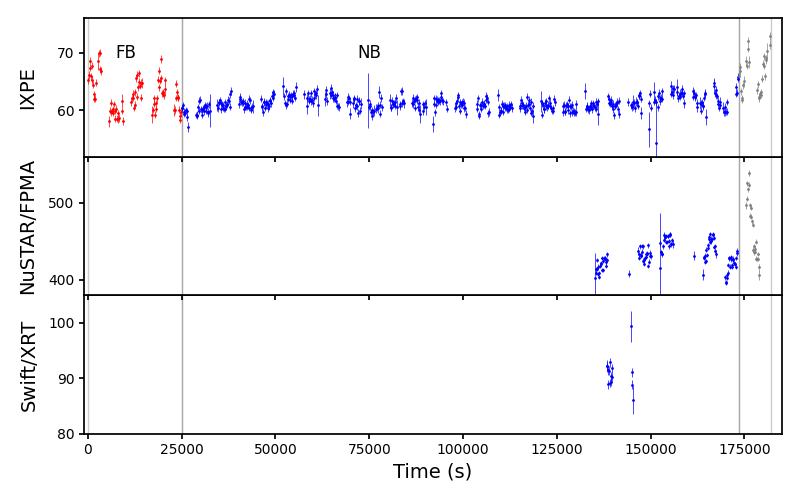}{0.5\textwidth}{}
              \fig{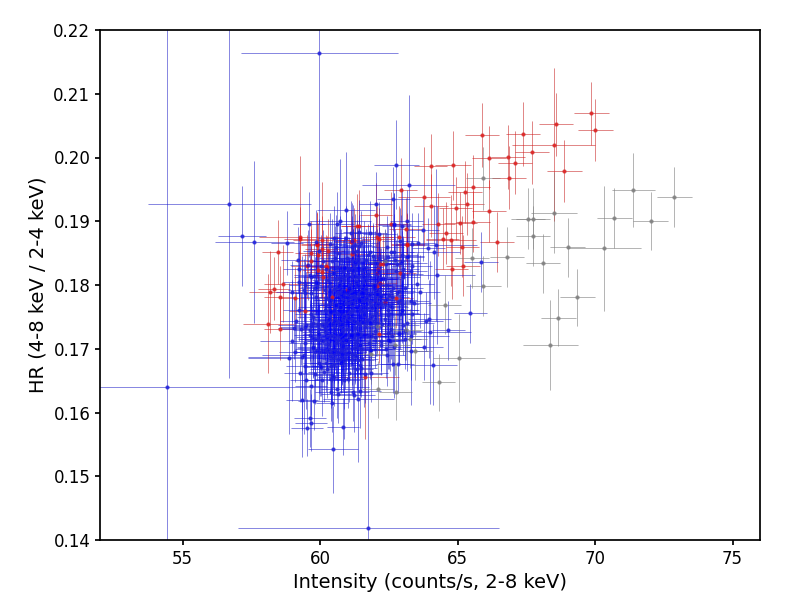}{0.42\textwidth}{}}    
    \caption{left panel: \textit{IXPE} (2-8 keV; combined DU1, Du2 and Du3), \textit{NuSTAR}/FPMA (3-79 keV) and \textit{Swift/XRT} (0.3-10 keV) lightcurve of GX 349+2. The normal branch and flaring branch are shown in red and blue. Right panel: Hardness intensity diagram (HID) for GX 349+2 using \textit{IXPE} observation. The blue and red colors show the NB and FB, respectively.}
    \label{lc}
\end{figure*}

\subsection{\textit{NuSTAR}}
The Nuclear Spectroscopic Telescope Array (\textit{NuSTAR}; \cite{Harrison2013}) is the first X-ray space observatory dedicated to studying astrophysical objects in high energy (3-79 keV). It consists of two co-aligned hard X-ray telescopes (FPMA and FPMB). The \textit{NuSTAR} observed the GX 349+2 on 2024 September 07 and 2024 September 08, simultaneously with \textit{IXPE}. An in-depth spectral analysis of \textit{NuSTAR} data is not within the scope of this article. So, we only used the \textit{NuSTAR} data on 2024 September 08 (obsid:\texttt{91002333004}), which exposure time is 10.1 ks. We used the \texttt{NuSTARDAS} pipeline \texttt{v0.4.9} and \texttt{CALDB v20241126} to reduce the \textit{NuSTAR} data. DS9 software was used to choose a circular source region with a radius of 120 arcsec at the source location and a circular region outside the source region with the same area for the background. Since the source is bright, we additionally used the \texttt{statusexpr="(STATUS==b0000xxx00xxxx000)\&\&\\(SHIELD==0)"} keyword in \texttt{nupipeline}. Then, the \texttt{NUPRODUCTS} tool was employed to extract spectra and light-curve files from both modules. Before spectral fitting, we rebin the source spectra for a minimum of 30 counts per bin using the \texttt{GRPPHA} task.

\section{Results} \label{sec:results}
\subsection{Lightcurve and HID}
The light curve of \textit{IXPE} (2-8 keV; combined all DUs), \textit{NuSTAR} (3-79 keV; FPMA only), and \textit{Swift/XRT} (0.3-10 keV) for GX 349+2 are shown in Figure \ref{lc}. Lightcurve obtained from \textit{IXPE} shows flux variation in that time. During the \textit{IXPE} observation, GX 349+2 transitioned from FB to NB and again back to FB (see Figure \ref{lc}).
\textit{NuSTAR} and \textit{Swift/XRT} also observed this source in the NB. We divided the data into FB and NB based on flux variation, as shown in Figure \ref{lc}. The data points in red and blue colors correspond to FB and NB, respectively. 
The hardness intensity diagram (HID) for GX 349+2 using \textit{IXPE} observation is shown in the right panel of Figure \ref{lc}. Here, the hardness ratio is the ratio between the hard band (4-8 keV) and the soft band (2-4 keV).


\begin{table*}[]
\centering
\caption{Spectral parameters of GX 349+2 during NB.}
\begin{tabular}{lcccccc}
\hline
Components & Parameters & Model-1 & Model-2 & Model-3 & Model-4 & Model-5\\ \hline
tbabs   & $n_\mathrm{H}$    & $0.83_{-0.01}^{+0.01}$ & $0.84_{-0.01}^{+0.01}$ & $0.83_{-0.01}^{+0.01}$ & $0.56_{-0.03}^{+0.03}$   & $0.56_{-0.03}^{+0.03}$  \\
diskbb  & $kT_\mathrm{in}$ & $1.89_{-0.01}^{+0.01}$ & $1.80_{-0.02}^{+0.02}$   & $1.74_{-0.04}^{+0.03}$ & -    & -             \\
        & $norm$   & $67.4_{-1.6}^{+1.6}$ & $80.2_{-2.4}^{+2.6}$ & $89.7_{-5.5}^{+6.6}$ & - & -                   \\ 
compTT & T0 & - & - & -                 & $0.50_{-0.02}^{+0.02}$ & $0.48_{-0.02}^{+0.02}$       \\
        & $kT_e$   & - & - & - & $2.70_{-0.02}^{+0.02}$  & $2.65_{-0.02}^{+0.02}$       \\ 
        & $\tau$ & -  & - & -          & $6.72_{-0.16}^{+0.16}$ & $6.96_{-0.16}^{+0.16}$       \\ 
        & $norm$   & - & - & - & $1.20_{-0.03}^{+0.03}$ & $1.21_{-0.03}^{+0.03}$       \\ 
bbodyrad& $kT$  &   $2.73_{-0.02}^{+0.02}$ &   $2.64_{-0.02}^{+0.02}$ & - &   $1.24_{-0.02}^{+0.02}$ &  -   \\
        & $norm$   & $6.9_{-0.4}^{+0.4}$ & $9.5_{-0.5}^{+0.5}$ & - & $201.6_{-17.4}^{+17.6}$ & -       \\ 
compbb  & $kT$     & -  & - & $2.50_{-0.06}^{+0.07}$ & -  & $1.21_{-0.02}^{+0.02}$     \\
        & $kT_e$   & - & - & $1.7_{-0.47}^{+0.56}$ & -  & $50^f$    \\
        & $\tau$   & - & - & $0.75_{-0.21}^{+0.17}$ & -  & $0.05_{-0.01}^{+0.01}$     \\
        & $norm$   & - & - & $20.4_{-4.6}^{+5.0}$ & -  & $240_{-23}^{+23}$    \\ 
Gauss   & $E$      & - & $6.64_{-0.03}^{+0.03}$ & $6.63_{-0.03}^{+0.03}$ & $6.65_{-0.03}^{+0.03}$ & $6.65_{-0.03}^{+0.03}$      \\
        & $\sigma$ & - & $0.38_{-0.04}^{+0.05}$ & $0.42_{-0.05}^{+0.06}$ & $0.33_{-0.04}^{+0.05}$  & $0.34_{-0.04}^{+0.05}$ \\
        & $norm(\times10^{-3})$  & - & $7.8_{-0.7}^{+0.8}$ & $8.7_{-0.9}^{+1.1}$ & $6.9_{-0.7}^{+0.8}$ & $7.1_{-0.7}^{+0.8}$ \\
        \hline
        & $\chi^2/dof$  & 2343/1503 & 1588/1500 & 1571/1498 &  1543/1498 & 1520/1497     \\\hline

\end{tabular}
\\
\label{table_flux}
\end{table*}

\begin{figure} [h]
    \centering
    \includegraphics[width=0.9\linewidth]{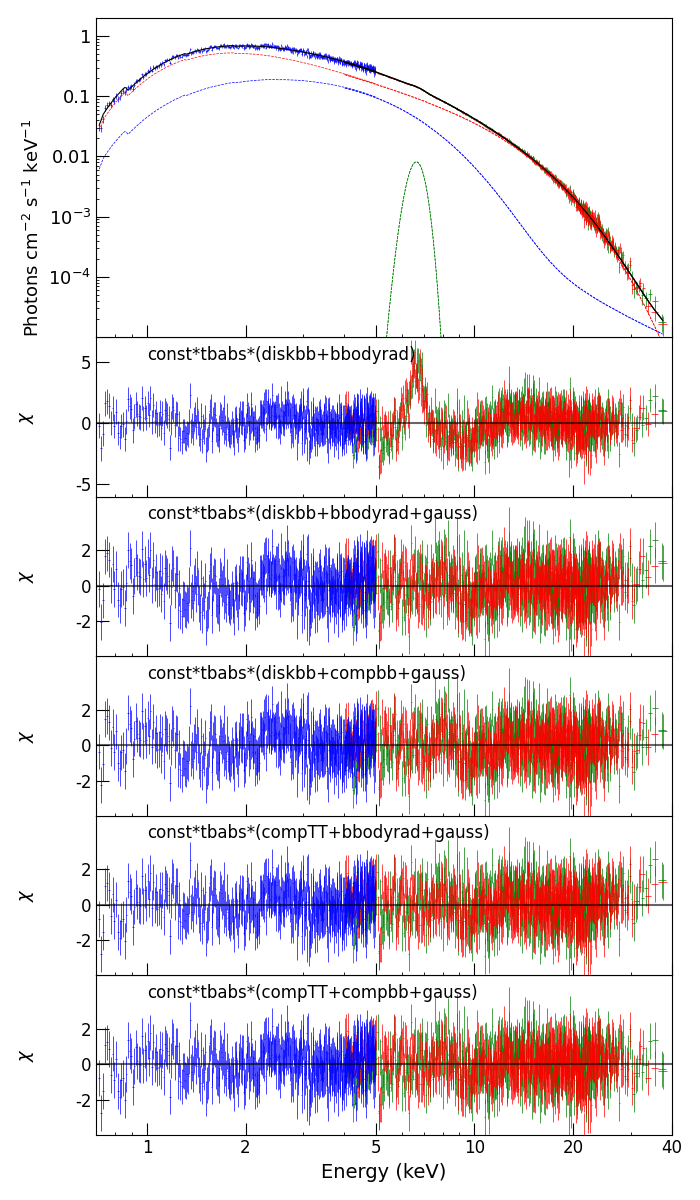}
    \caption{Spectra of GX 349+2, plotted with additive model components. The \textit{NuSTAR}/FPMA (4.0-40.0 keV), \textit{NuSTAR}/FPMA (4.0-40.0 keV), and \textit{Swift/XRT} (0.7-5.0 keV) data points are shown by green, red, and blue colors, respectively. The red, blue, and green dotted lines in the upper panel represent the \texttt{compTT}, \texttt{compbb}, and \texttt{Gaussian} components, respectively. The residual for each model employed in this study is provided, as indicated in the panel. Here, $\chi$ is (data-model)/error. }
    \label{spectra}
\end{figure}

\begin{table*}
  \centering
  \caption{Polarization of GX 349+2. The errors are at the 68 percent (1-sigma) confidence level.}\label{tab:pol_prop}
  \begin{tabular}{lccccc}
    \multicolumn{6}{c}{Model-independent polarization} \\\hline
    Location& Energy range (keV)     & MDP$_{99}$            & PD & PA & Detection significance\\ \hline
    FB      & 2.0-8.0   & 1.58 & $1.74\pm0.52$ & $19.4\pm8.9$ & $3.3\sigma$ \\\hline
            & 2.0-8.0   & 0.67 & $0.80\pm0.22$ & $35.4\pm7.9$ & $3.6\sigma$\\
    NB      & 2.0-4.0   & 0.63 & $0.78\pm0.22$ & $17.9\pm8.1$ & $3.6\sigma$\\
            & 4.0-8.0   & 1.21 & $1.32\pm0.40$ & $53.2\pm8.6$ & $3.3\sigma$\\\hline

  \end{tabular}
  \vspace{2\lineskip}
  \begin{tabular}{llcc}
    \multicolumn{4}{c}{Spectro-polarimetric analysis (NB)}\\\hline
    Case & Component &  PD (\%) & PA ($^\circ$) \\ \hline
    Full model polarized& - & $0.65\pm0.18$& $36.6\pm8.3$\\ \hline

    Individual components  & \texttt{compbb} & $4.42\pm1.83$ & $>67$ \\
    are polarized & \texttt{compTT} & $3.35\pm1.31$& $-1.4\pm11.5$ \\
                 & \texttt{Gaussian} & $<42$ & Unconstrained \\ \hline
    compbb is & \texttt{compTT} & $1.10\pm0.3$ & $32\pm9$\\
    unpolarized & \texttt{Gaussian} & $34\pm28$& $>48$\\ \hline

  \end{tabular}
\end{table*}

\begin{figure*} []
    \centering
    \gridline{\fig{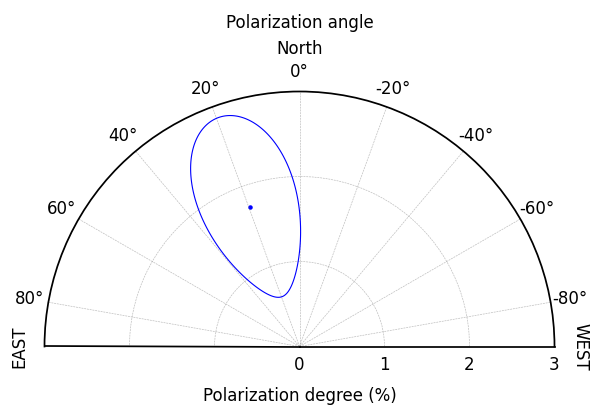}{0.5\textwidth}{}
              \fig{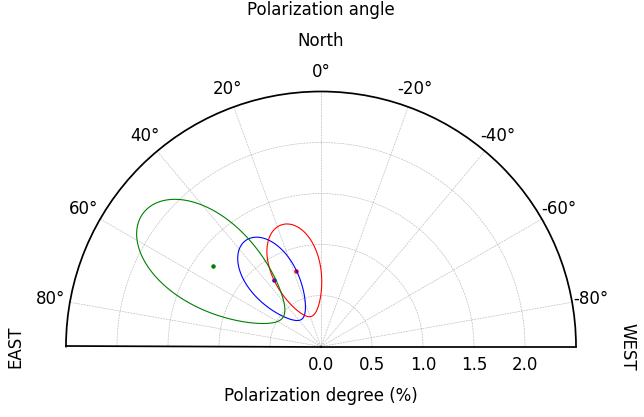}{0.5\textwidth}{}}    

    \caption{Left panel: The contour plot for FB in energy 2-8 keV. Right panel: The contour plot for NB. The blue, red, and green colors correspond to 2-8 keV, 2-4 keV, and 4-8 keV, respectively. Contours are at 90\% CL.}
    \label{contour}
\end{figure*}

\subsection{Spectroscopy of GX 349+2}
\label{spec_nu_xrt}
The joint \textit{Swift/XRT} (0.7-5.0 keV) and \textit{NuSTAR} (4.0-40.0 keV) spectrum during NB undergoes spectral analysis using \texttt{XSPEC} version \texttt{12.10.0h} \citep{Arnaud1996}. We started spectral fitting with the model (Model-1): \texttt{const*tbabs*(diskbb+bbodyrad)}. The energy-independent factor \texttt{const} aims to manage the calibration differences between \textit{Swift/XRT} and \textit{NuSTAR} FPMA/B. \texttt{tbabs} is an ISM absorption model, with cross-sections and abundances provided in \cite{Wilms2000}. \texttt{diskbb} is utilized to fit the disk blackbody component, whereas \texttt{bbodyrad} addresses the emission from the NS surface. 
The fitting of the data yielded a goodness of fit $\chi^2/\mathrm{dof}=2343/1503$. The residual shows an iron line feature at around 6.5 keV and take care by adding a \texttt{Gaussian} in the model (Model-2). The $\chi^2$ decreased by 755 for 3 degrees of freedom. The Gaussian fitted at $\sim6.64$ keV. Then, we replace the \texttt{bbodyrad} with the \texttt{compbb} model to check the spherical geometry. The \texttt{compbb} model takes care of the comptonized blackbody emission from the NS surface (Model-3) \citep{Nishimura1986}. The temperature of the blackbody ($kT$), the electron temperature of the corona ($kT_e$), and the optical depth of the corona ($\tau$) came out to be $\sim2.50\; \mathrm{keV}$, $\sim1.7\; \mathrm{keV}$, and $\sim0.75$, respectively. The $\chi^2/\mathrm{dof}$ comes out to be $1571/1498$. 

To check the slab geometry, we replaced the \texttt{diskbb} with \texttt{compTT} in Model-2. The \texttt{compTT} is an analytical model that describes Comptonization of soft photons in a hot plasma (Model-4) \citep{Titarchuk1994}.\texttt{compTT} contains the following parameters: input soft photon (Wien) temperature (T0), electron temperature of corona ($kT_e$), plasma optical depth ($\tau$), a parameter for geometry switch, and norm. The goodness of fit is $\chi^2/\mathrm{dof}=1543/1498$. Finally, we replaced \texttt{bbodyrad} with \texttt{compbb} in Model-4. The model (Model-5) becomes \texttt{const*tbabs*(compTT+compbb+Gaussian)}. 
Assuming the input seed photons from disk geometry for \texttt{compTT}, the fit achieved a goodness of fit $\chi^2/\mathrm{dof} = 1520/1497$. We can not constrain the $kT_e$ of \texttt{compbb}, and it tends to be a high temperature; therefore, we fixed it at 50 keV. The $n_\mathrm{H}$ came out to be $0.56\times10^{22}\;\mathrm{cm^{-2}}$. The input seed temperature for \texttt{compTT} and \texttt{compbb} was found to be $\sim0.48$ and $\sim1.21$ keV, respectively. The electron temperature of the corona for the \texttt{compTT} model came out to be $\sim2.65$ keV. 
The fitting process determined the optical depth of corona for \texttt{compTT} and \texttt{compbb} as $\sim6.96$ and $\sim0.05$, respectively. The estimated flux (0.7-40 keV) for GX 349+2 during NB was found to be $\sim2\times10^{-8}\;\mathrm{erg\;cm^{-2}\;s^{-1}}$.

\subsection{Polarization Measurements}
\subsubsection{Model-independent analysis}
The \texttt{PCUBE} algorithm is used to perform model-independent polarimetric analysis to estimate the polarization degree (PD) and polarization angle (PA) for the source during the FB and NB. The analysis is conducted using the calibration file (v.12). In the 2-8 keV energy range, the calculated PD for FB and NB are $\sim1.78$ and $\sim0.80$, respectively, while the PA for FB and NB are $\sim19.4$ and $\sim35.4$, respectively. 

Further, we estimate the energy-dependent PD and PA. We divided the data into 2-4 keV and 4-8 keV energy ranges. For FB, the value of PD obtained from the analysis is found to be less than MDP$_{99}$ in the \textit{IXPE} energy bands (2-4 keV and 4-8 keV). However, we detect significant polarization for NB ($>99$ \%) in these energy ranges. The PA and PD for NB in 2-4 keV are 0.78 and 17.9, respectively. The PA and PD for NB in 4-8 keV are 1.32 and 53.2, respectively. The PA and PD in FB and NB are displayed on a protractor plot in Figure \ref{contour}.

\subsubsection{Spectro-polarimetric analysis}
For spectro-polarimetric study, we fitted the \textit{I}, \textit{U}, and \textit{Q} pha files with the Model-5 mentioned in section \ref{spec_nu_xrt} and modified it with the \texttt{polconst} model. The model becomes \texttt{const*tbabs*polconst*(compTT+compbb+Gaussian)}. We fixed the $n_\mathrm{H}$ of \texttt{tbabs}, T0, $kT_\mathrm{e}$ of \texttt{compTT}, $kT$, $kT_e$, $\tau$ of \texttt{compbb}, and $E$, $\sigma$ of \texttt{Gauss} at the values estimated using simultaneous fitting of \textit{NuSTAR} and \textit{Swift/XRT}. We left the norm of \texttt{compTT}, \texttt{compbb}, and \texttt{Gaussian} free. Additionally, we also left $\tau$ of \texttt{compTT} free. We added a 1\% systematic error during fitting. This yields a $\chi^2$ of 455 for 442 dof. We estimated the PD and PA to be $\sim0.65$ and $\sim36.6$, respectively. A \texttt{polconst} model was then added for each of the \texttt{compTT}, \texttt{compbb}, and \texttt{Gaussian} components to examine their polarization properties separately. Furthermore, we investigated the case where \texttt{compbb} is not polarized by setting PD and PA equal to 0 for \texttt{compbb}. The PD and PA for every scenario are listed in Table \ref{tab:pol_prop}.

\section{Discussion and Conclusion} \label{sec:dis}
Research on coronal geometry in NSXRBs relied primarily on spectral studies, and there is dispute about the exact coronal geometry.
In this study, we used \textit{IXPE}, \textit{NuSTAR}, and \textit{Swift/XRT} data to probe the geometry of the X-ray emitting region through spectral and polarimetric studies.
The simultaneous data of \textit{NuSTAR} and \textit{Swift/XRT} during the NB is well modeled by \texttt{const*tbabs(compTT+compbb+Gauss)}.
The optical depth of \texttt{compbb} is $0.05$, which is less than one, indicating that there is negligible corona around the NS surface. The seed photon temperature of \texttt{compbb} is consistent with that of \texttt{bbodyrad} in Model-4. Further, spectral modeling suggests that the seed photons emitted from the disk with temperature $kT\sim0.48$ Comptonized in low electron temperature ($kT_\mathrm{e}\sim2.65\;\mathrm{keV}$) and high Thomson optical depth ($\tau\sim6.96$) corona. This paradigm aligns with the Western model. 
The \texttt{Gaussian} component in the model is the result of reprocessed emission from the accretion disk.

IXPE observed the GX 349+2 during FB and NB. In the 2-8 keV energy range, the PD reduced from $1.74\pm0.52$ to $0.80\pm0.22$, while the PA changed from $19.4\pm8.9$ to $35.4\pm7.9$ when the source transitioned from FB to NB. The observed PD for GX 349+2 during NB/FB is in the range of $0.6-2\%$, similar to other NSXRBs \citep{Ferinelli2023, Cocchi2011, Fabiani2024, LaMonaca2024, Rankin2024, Bhargava2024}. Additionally, we estimate the significant PA and PD for NB in the energy range of 2-4 keV and 4-8 keV. The PD for 2-4 and 4-8 keV are $0.78\pm0.22$ and $1.32\pm0.40$, respectively. The PA for 2-4 keV is $17.9\pm8.1$, while for 4-8 keV, it is $53.2\pm8.6$. PD and PA were both observed to increase with energy.

\cite{Gnarini2022} performed comprehensive polarimetric simulations of NS-LMXBs with two coronal configurations, namely, shell geometry and slab geometry. With different system inclinations and spectral states (HSS or LHS), the authors reported the expected PD and PA from such sources. 
The PD increases with energy in the case of slab geometry. The polarimetric studies of GX 349+2 during NB favor the slab geometry over the shell geometry. The possibility of alternative geometrical configurations cannot be ruled out, as simulations performed by \cite{Capitanio2023} for wedge geometry in addition to these two configurations showed the same trend for PD.

The spectra of GX 349+2 during NB consist of a blackbody component, a Comptonized component, and a reflection component. 
Each component contributes to the polarization properties of the system.
An upper limit of $1.5\%$ on PD from the spreading layer was proposed by \cite{Bobrikova2024}, and $0.5\%$ was determined from the boundary layer by \cite{Farinelli2024}. Several investigations have explored the polarization due to disk emission \citep{Dovciak2008, Li2009, Loktev2022}.
The PD measured using spectro-polarimetric studies is slightly lower than the model-independent PD, and negligible variation is observed in PA. The PD for \texttt{compbb} and \texttt{compTT} is $4.42\pm1.83\%$ and $3.35\pm1.31^\circ$, respectively. The polarization angle between \texttt{compbb} and \texttt{compTT} is $>60^\circ$. For the \texttt{Gaussian}, the upper limit of PD at $42\%$ is found, and PA is not well constrained. When we took the unpolarized emission from the NS surface, the PD and PA estimated for \texttt{compTT} are $1.10\pm0.30\%$ and $32\pm9^\circ$. The PD for \texttt{Gaussian} comes out to be $34\pm28\%$, and PA gives a lower limit of $48^\circ$. It is difficult to identify whether the observed PD is due to Comptonized emission or merely the reflection component \citep{LaMonaca2024x, Bhargava2024}. 

The spectral and polarimetric studies provide convincing evidence in favor of the slab geometry (western model) of GX 349+2 during NB. To better understand the corona geometry in FB, further \textit{IXPE} and \textit{XpoSat} observations are needed to determine the energy-dependent polarimetric properties of GX 349+2 during FB.

\begin{acknowledgments}
This research has made use of data and/or software provided by the High Energy Astrophysics Science Archive Research Center (HEASARC), which is a service of the Astrophysics Science Division at NASA/GSFC. This work reports observations obtained with the Imaging X-ray Polarimetry Explorer (\textit{IXPE}), a joint US (NASA) and Italian (ASI) mission, led by Marshall Space Flight Center (MSFC). The research uses data products provided by the \textit{IXPE} Science Operations Center (MSFC), using algorithms developed by the \textit{IXPE} Collaboration, and distributed by the High-Energy Astrophysics Science Archive Research Center (HEASARC). This research has made use of data from the \textit{NuSTAR} mission, a project led by the California Institute of Technology, managed by the Jet Propulsion Laboratory, and funded by the National Aeronautics and Space Administration. Data analysis was performed using the \textit{NuSTAR} Data Analysis Software (NuSTARDAS), jointly developed by the ASI Science Data Center (SSDC, Italy) and the California Institute of Technology (USA). We acknowledge the use of public data from the Swift data archive. 
\end{acknowledgments}



\facilities{\textit{IXPE}, \textit{NuSTAR}, Swift(XRT)}

\software{astropy \citep{2013A&A...558A..33A,2018AJ....156..123A,2022ApJ...935..167A}, matplotlib \citep{Hunter2007}, ixpeobssim \citep{Baldini2022}, DS9, HEASOFT 
          }


\bibliography{gx349}{}

\begin{thebibliography}{}
\expandafter\ifx\csname natexlab\endcsname\relax\def\natexlab#1{#1}\fi
\providecommand{\url}[1]{\href{#1}{#1}}
\providecommand{\dodoi}[1]{doi:~\href{http://doi.org/#1}{\nolinkurl{#1}}}
\providecommand{\doeprint}[1]{\href{http://ascl.net/#1}{\nolinkurl{http://ascl.net/#1}}}
\providecommand{\doarXiv}[1]{\href{https://arxiv.org/abs/#1}{\nolinkurl{https://arxiv.org/abs/#1}}}

\bibitem[{K.~A. {Arnaud}(1996){Arnaud}}]{Arnaud1996}
{Arnaud}, K.~A. 1996, in Astronomical Society of the Pacific Conference Series, Vol. 101, Astronomical Data Analysis Software and Systems V, ed. G.~H. {Jacoby} \& J.~{Barnes}, 17

\bibitem[{ {Astropy Collaboration} {et~al.}(2013){Astropy Collaboration}, {Robitaille}, {Tollerud}, {Greenfield}, {Droettboom}, {Bray}, {Aldcroft}, {Davis}, {Ginsburg}, {Price-Whelan}, {Kerzendorf}, {Conley}, {Crighton}, {Barbary}, {Muna}, {Ferguson}, {Grollier}, {Parikh}, {Nair}, {Unther}, {Deil}, {Woillez}, {Conseil}, {Kramer}, {Turner}, {Singer}, {Fox}, {Weaver}, {Zabalza}, {Edwards}, {Azalee Bostroem}, {Burke}, {Casey}, {Crawford}, {Dencheva}, {Ely}, {Jenness}, {Labrie}, {Lim}, {Pierfederici}, {Pontzen}, {Ptak}, {Refsdal}, {Servillat}, \& {Streicher}}]{2013A&A...558A..33A}
{Astropy Collaboration}, {Robitaille}, T.~P., {Tollerud}, E.~J., {et~al.} 2013, \bibinfo{title}{{Astropy: A community Python package for astronomy},} \aap, 558, A33, \dodoi{10.1051/0004-6361/201322068}

\bibitem[{ {Astropy Collaboration} {et~al.}(2018){Astropy Collaboration}, {Price-Whelan}, {Sip{\H{o}}cz}, {G{\"u}nther}, {Lim}, {Crawford}, {Conseil}, {Shupe}, {Craig}, {Dencheva}, {Ginsburg}, {VanderPlas}, {Bradley}, {P{\'e}rez-Su{\'a}rez}, {de Val-Borro}, {Aldcroft}, {Cruz}, {Robitaille}, {Tollerud}, {Ardelean}, {Babej}, {Bach}, {Bachetti}, {Bakanov}, {Bamford}, {Barentsen}, {Barmby}, {Baumbach}, {Berry}, {Biscani}, {Boquien}, {Bostroem}, {Bouma}, {Brammer}, {Bray}, {Breytenbach}, {Buddelmeijer}, {Burke}, {Calderone}, {Cano Rodr{\'\i}guez}, {Cara}, {Cardoso}, {Cheedella}, {Copin}, {Corrales}, {Crichton}, {D'Avella}, {Deil}, {Depagne}, {Dietrich}, {Donath}, {Droettboom}, {Earl}, {Erben}, {Fabbro}, {Ferreira}, {Finethy}, {Fox}, {Garrison}, {Gibbons}, {Goldstein}, {Gommers}, {Greco}, {Greenfield}, {Groener}, {Grollier}, {Hagen}, {Hirst}, {Homeier}, {Horton}, {Hosseinzadeh}, {Hu}, {Hunkeler}, {Ivezi{\'c}}, {Jain}, {Jenness}, {Kanarek}, {Kendrew}, {Kern}, {Kerzendorf}, {Khvalko}, {King}, {Kirkby}, {Kulkarni},
  {Kumar}, {Lee}, {Lenz}, {Littlefair}, {Ma}, {Macleod}, {Mastropietro}, {McCully}, {Montagnac}, {Morris}, {Mueller}, {Mumford}, {Muna}, {Murphy}, {Nelson}, {Nguyen}, {Ninan}, {N{\"o}the}, {Ogaz}, {Oh}, {Parejko}, {Parley}, {Pascual}, {Patil}, {Patil}, {Plunkett}, {Prochaska}, {Rastogi}, {Reddy Janga}, {Sabater}, {Sakurikar}, {Seifert}, {Sherbert}, {Sherwood-Taylor}, {Shih}, {Sick}, {Silbiger}, {Singanamalla}, {Singer}, {Sladen}, {Sooley}, {Sornarajah}, {Streicher}, {Teuben}, {Thomas}, {Tremblay}, {Turner}, {Terr{\'o}n}, {van Kerkwijk}, {de la Vega}, {Watkins}, {Weaver}, {Whitmore}, {Woillez}, {Zabalza}, \& {Astropy Contributors}}]{2018AJ....156..123A}
{Astropy Collaboration}, {Price-Whelan}, A.~M., {Sip{\H{o}}cz}, B.~M., {et~al.} 2018, \bibinfo{title}{{The Astropy Project: Building an Open-science Project and Status of the v2.0 Core Package},} \aj, 156, 123, \dodoi{10.3847/1538-3881/aabc4f}

\bibitem[{ {Astropy Collaboration} {et~al.}(2022){Astropy Collaboration}, {Price-Whelan}, {Lim}, {Earl}, {Starkman}, {Bradley}, {Shupe}, {Patil}, {Corrales}, {Brasseur}, {N{\"o}the}, {Donath}, {Tollerud}, {Morris}, {Ginsburg}, {Vaher}, {Weaver}, {Tocknell}, {Jamieson}, {van Kerkwijk}, {Robitaille}, {Merry}, {Bachetti}, {G{\"u}nther}, {Aldcroft}, {Alvarado-Montes}, {Archibald}, {B{\'o}di}, {Bapat}, {Barentsen}, {Baz{\'a}n}, {Biswas}, {Boquien}, {Burke}, {Cara}, {Cara}, {Conroy}, {Conseil}, {Craig}, {Cross}, {Cruz}, {D'Eugenio}, {Dencheva}, {Devillepoix}, {Dietrich}, {Eigenbrot}, {Erben}, {Ferreira}, {Foreman-Mackey}, {Fox}, {Freij}, {Garg}, {Geda}, {Glattly}, {Gondhalekar}, {Gordon}, {Grant}, {Greenfield}, {Groener}, {Guest}, {Gurovich}, {Handberg}, {Hart}, {Hatfield-Dodds}, {Homeier}, {Hosseinzadeh}, {Jenness}, {Jones}, {Joseph}, {Kalmbach}, {Karamehmetoglu}, {Ka{\l}uszy{\'n}ski}, {Kelley}, {Kern}, {Kerzendorf}, {Koch}, {Kulumani}, {Lee}, {Ly}, {Ma}, {MacBride}, {Maljaars}, {Muna}, {Murphy}, {Norman},
  {O'Steen}, {Oman}, {Pacifici}, {Pascual}, {Pascual-Granado}, {Patil}, {Perren}, {Pickering}, {Rastogi}, {Roulston}, {Ryan}, {Rykoff}, {Sabater}, {Sakurikar}, {Salgado}, {Sanghi}, {Saunders}, {Savchenko}, {Schwardt}, {Seifert-Eckert}, {Shih}, {Jain}, {Shukla}, {Sick}, {Simpson}, {Singanamalla}, {Singer}, {Singhal}, {Sinha}, {Sip{\H{o}}cz}, {Spitler}, {Stansby}, {Streicher}, {{\v{S}}umak}, {Swinbank}, {Taranu}, {Tewary}, {Tremblay}, {de Val-Borro}, {Van Kooten}, {Vasovi{\'c}}, {Verma}, {de Miranda Cardoso}, {Williams}, {Wilson}, {Winkel}, {Wood-Vasey}, {Xue}, {Yoachim}, {Zhang}, {Zonca}, \& {Astropy Project Contributors}}]{2022ApJ...935..167A}
{Astropy Collaboration}, {Price-Whelan}, A.~M., {Lim}, P.~L., {et~al.} 2022, \bibinfo{title}{{The Astropy Project: Sustaining and Growing a Community-oriented Open-source Project and the Latest Major Release (v5.0) of the Core Package},} \apj, 935, 167, \dodoi{10.3847/1538-4357/ac7c74}

\bibitem[{L. {Baldini} {et~al.}(2022){Baldini}, {Bucciantini}, {Lalla}, {Ehlert}, {Manfreda}, {Negro}, {Omodei}, {Pesce-Rollins}, {Sgr{\`o}}, \& {Silvestri}}]{Baldini2022}
{Baldini}, L., {Bucciantini}, N., {Lalla}, N.~D., {et~al.} 2022, \bibinfo{title}{{ixpeobssim: A simulation and analysis framework for the imaging X-ray polarimetry explorer},} SoftwareX, 19, 101194, \dodoi{10.1016/j.softx.2022.101194}

\bibitem[{D. {Barret}(2001){Barret}}]{Barret2001}
{Barret}, D. 2001, \bibinfo{title}{{The broad band x-ray/hard x-ray spectra of accreting neutron stars},} Advances in Space Research, 28, 307, \dodoi{10.1016/S0273-1177(01)00414-8}

\bibitem[{Y. {Bhargava} {et~al.}(2024){Bhargava}, {Russell}, {Ng}, {Balasubramanian}, {Zhang}, {Ravi}, {Jadoliya}, {Bhattacharyya}, {Pahari}, {Homan}, {Marshall}, {Chakrabarty}, {Carotenuto}, \& {Kaushik}}]{Bhargava2024}
{Bhargava}, Y., {Russell}, T.~D., {Ng}, M., {et~al.} 2024, \bibinfo{title}{{X-ray and Radio Campaign of the Z-source GX 340+0 II: the X-ray polarization in the normal branch},} arXiv e-prints, arXiv:2411.00350, \dodoi{10.48550/arXiv.2411.00350}

\bibitem[{A. {Bobrikova} {et~al.}(2024){Bobrikova}, {Poutanen}, \& {Loktev}}]{Bobrikova2024}
{Bobrikova}, A., {Poutanen}, J., \& {Loktev}, V. 2024, \bibinfo{title}{{Polarized radiation coming from the spreading layer of the weakly magnetized neutron stars},} arXiv e-prints, arXiv:2409.16023, \dodoi{10.48550/arXiv.2409.16023}

\bibitem[{D.~N. {Burrows} {et~al.}(2005){Burrows}, {Hill}, {Nousek}, {Kennea}, {Wells}, {Osborne}, {Abbey}, {Beardmore}, {Mukerjee}, {Short}, {Chincarini}, {Campana}, {Citterio}, {Moretti}, {Pagani}, {Tagliaferri}, {Giommi}, {Capalbi}, {Tamburelli}, {Angelini}, {Cusumano}, {Br{\"a}uninger}, {Burkert}, \& {Hartner}}]{Burrows2005}
{Burrows}, D.~N., {Hill}, J.~E., {Nousek}, J.~A., {et~al.} 2005, \bibinfo{title}{{The Swift X-Ray Telescope},} \ssr, 120, 165, \dodoi{10.1007/s11214-005-5097-2}

\bibitem[{E.~M. {Cackett} {et~al.}(2008){Cackett}, {Miller}, {Bhattacharyya}, {Grindlay}, {Homan}, {van der Klis}, {Miller}, {Strohmayer}, \& {Wijnands}}]{Cackett2008}
{Cackett}, E.~M., {Miller}, J.~M., {Bhattacharyya}, S., {et~al.} 2008, \bibinfo{title}{{Relativistic Iron Emission Lines in Neutron Star Low-Mass X-Ray Binaries as Probes of Neutron Star Radii},} \apj, 674, 415, \dodoi{10.1086/524936}

\bibitem[{F. {Capitanio} {et~al.}(2023){Capitanio}, {Fabiani}, {Gnarini}, {Ursini}, {Ferrigno}, {Matt}, {Poutanen}, {Cocchi}, {Mikusincova}, {Farinelli}, {Bianchi}, {Kajava}, {Muleri}, {Sanchez-Fernandez}, {Soffitta}, {Wu}, {Agudo}, {Antonelli}, {Bachetti}, {Baldini}, {Baumgartner}, {Bellazzini}, {Bongiorno}, {Bonino}, {Brez}, {Bucciantini}, {Castellano}, {Cavazzuti}, {Ciprini}, {Costa}, {De Rosa}, {Del Monte}, {Di Gesu}, {Di Lalla}, {Di Marco}, {Donnarumma}, {Doroshenko}, {Dov{\v{c}}iak}, {Ehlert}, {Enoto}, {Evangelista}, {Ferrazzoli}, {Garcia}, {Gunji}, {Hayashida}, {Heyl}, {Iwakiri}, {Jorstad}, {Karas}, {Kitaguchi}, {Kolodziejczak}, {Krawczynski}, {La Monaca}, {Latronico}, {Liodakis}, {Maldera}, {Manfreda}, {Marin}, {Marinucci}, {Marscher}, {Marshall}, {Mitsuishi}, {Mizuno}, {Ng}, {O'Dell}, {Omodei}, {Oppedisano}, {Papitto}, {Pavlov}, {Peirson}, {Perri}, {Pesce-Rollins}, {Petrucci}, {Pilia}, {Possenti}, {Puccetti}, {Ramsey}, {Rankin}, {Ratheesh}, {Romani}, {Sgr{\`o}}, {Slane}, {Spandre}, {Tamagawa},
  {Tavecchio}, {Taverna}, {Tawara}, {Tennant}, {Thomas}, {Tombesi}, {Trois}, {Tsygankov}, {Turolla}, {Vink}, {Weisskopf}, {Xie}, \& {Zane}}]{Capitanio2023}
{Capitanio}, F., {Fabiani}, S., {Gnarini}, A., {et~al.} 2023, \bibinfo{title}{{Polarization Properties of the Weakly Magnetized Neutron Star X-Ray Binary GS 1826-238 in the High Soft State},} \apj, 943, 129, \dodoi{10.3847/1538-4357/acae88}

\bibitem[{D.~J. {Christian} \& J.~H. {Swank}(1997){Christian} \& {Swank}}]{Christian1997}
{Christian}, D.~J., \& {Swank}, J.~H. 1997, \bibinfo{title}{{The Survey of Low-Mass X-Ray Binaries with the Einstein Observatory Solid-State Spectrometer and Monitor Proportional Counter},} \apjs, 109, 177, \dodoi{10.1086/312970}

\bibitem[{M.~J. {Church} {et~al.}(2012){Church}, {Gibiec}, {Ba{\l}uci{\'n}ska-Church}, \& {Jackson}}]{Church2012}
{Church}, M.~J., {Gibiec}, A., {Ba{\l}uci{\'n}ska-Church}, M., \& {Jackson}, N.~K. 2012, \bibinfo{title}{{Spectral investigations of the nature of the Scorpius X-1 like sources},} \aap, 546, A35, \dodoi{10.1051/0004-6361/201218987}

\bibitem[{M. {Cocchi} {et~al.}(2011){Cocchi}, {Farinelli}, \& {Paizis}}]{Cocchi2011}
{Cocchi}, M., {Farinelli}, R., \& {Paizis}, A. 2011, \bibinfo{title}{{BeppoSAX view of the NS-LMXB GS 1826-238},} \aap, 529, A155, \dodoi{10.1051/0004-6361/201016241}

\bibitem[{M. {Cocchi} {et~al.}(2023){Cocchi}, {Gnarini}, {Fabiani}, {Ursini}, {Poutanen}, {Capitanio}, {Bobrikova}, {Farinelli}, {Paizis}, {Sidoli}, {Veledina}, {Bianchi}, {Di Marco}, {Ingram}, {Kajava}, {La Monaca}, {Matt}, {Malacaria}, {Miku{\v{s}}incov{\'a}}, {Rankin}, {Zane}, {Agudo}, {Antonelli}, {Bachetti}, {Baldini}, {Baumgartner}, {Bellazzini}, {Bongiorno}, {Bonino}, {Brez}, {Bucciantini}, {Castellano}, {Cavazzuti}, {Chen}, {Ciprini}, {Costa}, {De Rosa}, {Del Monte}, {Di Gesu}, {Di Lalla}, {Donnarumma}, {Doroshenko}, {Dov{\v{c}}iak}, {Ehlert}, {Enoto}, {Evangelista}, {Ferrazzoli}, {Garcia}, {Gunji}, {Hayashida}, {Heyl}, {Iwakiri}, {Jorstad}, {Kaaret}, {Karas}, {Kislat}, {Kitaguchi}, {Kolodziejczak}, {Krawczynski}, {Latronico}, {Liodakis}, {Maldera}, {Manfreda}, {Marin}, {Marinucci}, {Marscher}, {Marshall}, {Massaro}, {Mitsuishi}, {Mizuno}, {Muleri}, {Negro}, {Ng}, {O'Dell}, {Omodei}, {Oppedisano}, {Papitto}, {Pavlov}, {Peirson}, {Perri}, {Pesce-Rollins}, {Petrucci}, {Pilia}, {Possenti}, {Puccetti},
  {Ramsey}, {Ratheesh}, {Roberts}, {Romani}, {Sgr{\`o}}, {Slane}, {Soffitta}, {Spandre}, {Swartz}, {Tamagawa}, {Tavecchio}, {Taverna}, {Tawara}, {Tennant}, {Thomas}, {Tombesi}, {Trois}, {Tsygankov}, {Turolla}, {Vink}, {Weisskopf}, {Wu}, \& {Xie}}]{Cocchi2023}
{Cocchi}, M., {Gnarini}, A., {Fabiani}, S., {et~al.} 2023, \bibinfo{title}{{Discovery of strongly variable X-ray polarization in the neutron star low-mass X-ray binary transient XTE J1701{\ensuremath{-}}462},} \aap, 674, L10, \dodoi{10.1051/0004-6361/202346275}

\bibitem[{B.~A. {Cooke} \& T.~J. {Ponman}(1991){Cooke} \& {Ponman}}]{Cooke1991}
{Cooke}, B.~A., \& {Ponman}, T.~J. 1991, \bibinfo{title}{{A variable radio source in the error box of the bright galactic X-ray source GX 349+2.},} \aap, 244, 358

\bibitem[{B.~M. {Coughenour} {et~al.}(2018){Coughenour}, {Cackett}, {Miller}, \& {Ludlam}}]{Coughenour2018}
{Coughenour}, B.~M., {Cackett}, E.~M., {Miller}, J.~M., \& {Ludlam}, R.~M. 2018, \bibinfo{title}{{A NuSTAR Observation of the Low-mass X-Ray Binary GX 349+2 throughout the Z-track},} \apj, 867, 64, \dodoi{10.3847/1538-4357/aae098}

\bibitem[{N. {Degenaar} {et~al.}(2018){Degenaar}, {Ballantyne}, {Belloni}, {Chakraborty}, {Chen}, {Ji}, {Kretschmar}, {Kuulkers}, {Li}, {Maccarone}, {Malzac}, {Zhang}, \& {Zhang}}]{Degenaar2018}
{Degenaar}, N., {Ballantyne}, D.~R., {Belloni}, T., {et~al.} 2018, \bibinfo{title}{{Accretion Disks and Coronae in the X-Ray Flashlight},} \ssr, 214, 15, \dodoi{10.1007/s11214-017-0448-3}

\bibitem[{A. {Di Marco} {et~al.}(2023){Di Marco}, {Soffitta}, {Costa}, {Ferrazzoli}, {La Monaca}, {Rankin}, {Ratheesh}, {Xie}, {Baldini}, {Del Monte}, {Ehlert}, {Fabiani}, {Kim}, {Muleri}, {O'Dell}, {Ramsey}, {Rubini}, {Sgr{\`o}}, {Silvestri}, {Tennant}, \& {Weisskopf}}]{Dimarco2023}
{Di Marco}, A., {Soffitta}, P., {Costa}, E., {et~al.} 2023, \bibinfo{title}{{Handling the Background in IXPE Polarimetric Data},} \aj, 165, 143, \dodoi{10.3847/1538-3881/acba0f}

\bibitem[{T. {Di Salvo} {et~al.}(2001){Di Salvo}, {Robba}, {Iaria}, {Stella}, {Burderi}, \& {Israel}}]{Disalvo2001}
{Di Salvo}, T., {Robba}, N.~R., {Iaria}, R., {et~al.} 2001, \bibinfo{title}{{Detection of a Hard Tail in the X-Ray Spectrum of the Z Source GX 349+2},} \apj, 554, 49, \dodoi{10.1086/321353}

\bibitem[{G.~Q. {Ding} {et~al.}(2016){Ding}, {Zhang}, {Wang}, {Li}, {Qu}, \& {Huang}}]{Ding2016}
{Ding}, G.~Q., {Zhang}, W.~Y., {Wang}, Y.~N., {et~al.} 2016, \bibinfo{title}{{The cross-correlation analysis in Z source GX 349+2},} \mnras, 455, 2959, \dodoi{10.1093/mnras/stv2459}

\bibitem[{M. {Dov{\v{c}}iak} {et~al.}(2008){Dov{\v{c}}iak}, {Muleri}, {Goosmann}, {Karas}, \& {Matt}}]{Dovciak2008}
{Dov{\v{c}}iak}, M., {Muleri}, F., {Goosmann}, R.~W., {Karas}, V., \& {Matt}, G. 2008, \bibinfo{title}{{Thermal disc emission from a rotating black hole: X-ray polarization signatures},} \mnras, 391, 32, \dodoi{10.1111/j.1365-2966.2008.13872.x}

\bibitem[{S. {Fabiani} {et~al.}(2024){Fabiani}, {Capitanio}, {Iaria}, {Poutanen}, {Gnarini}, {Ursini}, {Farinelli}, {Bobrikova}, {Steiner}, {Svoboda}, {Anitra}, {Baglio}, {Carotenuto}, {Del Santo}, {Ferrigno}, {Lewis}, {Russell}, {Russell}, {van den Eijnden}, {Cocchi}, {Di Marco}, {La Monaca}, {Liu}, {Rankin}, {Weisskopf}, {Xie}, {Bianchi}, {Burderi}, {Di Salvo}, {Egron}, {Illiano}, {Kaaret}, {Matt}, {Miku{\v{s}}incov{\'a}}, {Muleri}, {Papitto}, {Agudo}, {Antonelli}, {Bachetti}, {Baldini}, {Baumgartner}, {Bellazzini}, {Bongiorno}, {Bonino}, {Brez}, {Bucciantini}, {Castellano}, {Cavazzuti}, {Chen}, {Ciprini}, {Costa}, {De Rosa}, {Del Monte}, {Di Gesu}, {Di Lalla}, {Donnarumma}, {Doroshenko}, {Dov{\v{c}}iak}, {Ehlert}, {Enoto}, {Evangelista}, {Ferrazzoli}, {Garcia}, {Gunji}, {Hayashida}, {Heyl}, {Iwakiri}, {Jorstad}, {Karas}, {Kislat}, {Kitaguchi}, {Kolodziejczak}, {Krawczynski}, {Latronico}, {Liodakis}, {Maldera}, {Manfreda}, {Marin}, {Marinucci}, {Marscher}, {Marshall}, {Massaro}, {Mitsuishi}, {Mizuno},
  {Negro}, {Ng}, {O'Dell}, {Omodei}, {Oppedisano}, {Pavlov}, {Peirson}, {Perri}, {Pesce-Rollins}, {Petrucci}, {Pilia}, {Possenti}, {Puccetti}, {Ramsey}, {Ratheesh}, {Roberts}, {Romani}, {Sgr{\`o}}, {Slane}, {Soffitta}, {Spandre}, {Swartz}, {Tamagawa}, {Tavecchio}, {Taverna}, {Tawara}, {Tennant}, {Thomas}, {Tombesi}, {Trois}, {Tsygankov}, {Turolla}, {Vink}, {Wu}, \& {Zane}}]{Fabiani2024}
{Fabiani}, S., {Capitanio}, F., {Iaria}, R., {et~al.} 2024, \bibinfo{title}{{Discovery of a variable energy-dependent X-ray polarization in the accreting neutron star GX 5{\ensuremath{-}}1},} \aap, 684, A137, \dodoi{10.1051/0004-6361/202347374}

\bibitem[{R. {Farinelli} {et~al.}(2024){Farinelli}, {Waghmare}, {Ducci}, \& {Santangelo}}]{Farinelli2024}
{Farinelli}, R., {Waghmare}, A., {Ducci}, L., \& {Santangelo}, A. 2024, \bibinfo{title}{{The polarization of the boundary layer around weakly magnetized neutron stars in X-ray binaries},} \aap, 684, A62, \dodoi{10.1051/0004-6361/202348915}

\bibitem[{R. {Farinelli} {et~al.}(2023){Farinelli}, {Fabiani}, {Poutanen}, {Ursini}, {Ferrigno}, {Bianchi}, {Cocchi}, {Capitanio}, {De Rosa}, {Gnarini}, {Kislat}, {Matt}, {Mikusincova}, {Muleri}, {Agudo}, {Antonelli}, {Bachetti}, {Baldini}, {Baumgartner}, {Bellazzini}, {Bongiorno}, {Bonino}, {Brez}, {Bucciantini}, {Castellano}, {Cavazzuti}, {Ciprini}, {Costa}, {Del Monte}, {Di Gesu}, {Di Lalla}, {Di Marco}, {Donnarumma}, {Doroshenko}, {Dov{\v{c}}iak}, {Ehlert}, {Enoto}, {Evangelista}, {Ferrazzoli}, {Garcia}, {Gunji}, {Hayashida}, {Heyl}, {Iwakiri}, {Jorstad}, {Karas}, {Kitaguchi}, {Kolodziejczak}, {Krawczynski}, {La Monaca}, {Latronico}, {Liodakis}, {Maldera}, {Manfreda}, {Marin}, {Marscher}, {Marshall}, {Mitsuishi}, {Mizuno}, {Ng}, {O'Dell}, {Omodei}, {Oppedisano}, {Papitto}, {Pavlov}, {Peirson}, {Perri}, {Pesce-Rollins}, {Petrucci}, {Pilia}, {Possenti}, {Puccetti}, {Ramsey}, {Rankin}, {Ratheesh}, {Romani}, {Sgr{\`o}}, {Slane}, {Soffitta}, {Spandre}, {Tamagawa}, {Tavecchio}, {Taverna}, {Tawara}, {Tennant},
  {Thomas}, {Tombesi}, {Trois}, {Tsygankov}, {Turolla}, {Vink}, {Weisskopf}, {Wu}, {Xie}, \& {Zane}}]{Ferinelli2023}
{Farinelli}, R., {Fabiani}, S., {Poutanen}, J., {et~al.} 2023, \bibinfo{title}{{Accretion geometry of the neutron star low mass X-ray binary Cyg X-2 from X-ray polarization measurements},} \mnras, 519, 3681, \dodoi{10.1093/mnras/stac3726}

\bibitem[{N. {Gehrels} {et~al.}(2004){Gehrels}, {Chincarini}, {Giommi}, {Mason}, {Nousek}, {Wells}, {White}, {Barthelmy}, {Burrows}, {Cominsky}, {Hurley}, {Marshall}, {M{\'e}sz{\'a}ros}, {Roming}, {Angelini}, {Barbier}, {Belloni}, {Campana}, {Caraveo}, {Chester}, {Citterio}, {Cline}, {Cropper}, {Cummings}, {Dean}, {Feigelson}, {Fenimore}, {Frail}, {Fruchter}, {Garmire}, {Gendreau}, {Ghisellini}, {Greiner}, {Hill}, {Hunsberger}, {Krimm}, {Kulkarni}, {Kumar}, {Lebrun}, {Lloyd-Ronning}, {Markwardt}, {Mattson}, {Mushotzky}, {Norris}, {Osborne}, {Paczynski}, {Palmer}, {Park}, {Parsons}, {Paul}, {Rees}, {Reynolds}, {Rhoads}, {Sasseen}, {Schaefer}, {Short}, {Smale}, {Smith}, {Stella}, {Tagliaferri}, {Takahashi}, {Tashiro}, {Townsley}, {Tueller}, {Turner}, {Vietri}, {Voges}, {Ward}, {Willingale}, {Zerbi}, \& {Zhang}}]{Gehrels2004}
{Gehrels}, N., {Chincarini}, G., {Giommi}, P., {et~al.} 2004, \bibinfo{title}{{The Swift Gamma-Ray Burst Mission},} \apj, 611, 1005, \dodoi{10.1086/422091}

\bibitem[{A. {Gnarini} {et~al.}(2022){Gnarini}, {Ursini}, {Matt}, {Bianchi}, {Capitanio}, {Cocchi}, {Farinelli}, \& {Zhang}}]{Gnarini2022}
{Gnarini}, A., {Ursini}, F., {Matt}, G., {et~al.} 2022, \bibinfo{title}{{Polarization properties of weakly magnetized neutron stars in low-mass X-ray binaries},} \mnras, 514, 2561, \dodoi{10.1093/mnras/stac1523}

\bibitem[{F.~A. {Harrison} {et~al.}(2013){Harrison}, {Craig}, {Christensen}, {Hailey}, {Zhang}, {Boggs}, {Stern}, {Cook}, {Forster}, {Giommi}, {Grefenstette}, {Kim}, {Kitaguchi}, {Koglin}, {Madsen}, {Mao}, {Miyasaka}, {Mori}, {Perri}, {Pivovaroff}, {Puccetti}, {Rana}, {Westergaard}, {Willis}, {Zoglauer}, {An}, {Bachetti}, {Barri{\`e}re}, {Bellm}, {Bhalerao}, {Brejnholt}, {Fuerst}, {Liebe}, {Markwardt}, {Nynka}, {Vogel}, {Walton}, {Wik}, {Alexander}, {Cominsky}, {Hornschemeier}, {Hornstrup}, {Kaspi}, {Madejski}, {Matt}, {Molendi}, {Smith}, {Tomsick}, {Ajello}, {Ballantyne}, {Balokovi{\'c}}, {Barret}, {Bauer}, {Blandford}, {Brandt}, {Brenneman}, {Chiang}, {Chakrabarty}, {Chenevez}, {Comastri}, {Dufour}, {Elvis}, {Fabian}, {Farrah}, {Fryer}, {Gotthelf}, {Grindlay}, {Helfand}, {Krivonos}, {Meier}, {Miller}, {Natalucci}, {Ogle}, {Ofek}, {Ptak}, {Reynolds}, {Rigby}, {Tagliaferri}, {Thorsett}, {Treister}, \& {Urry}}]{Harrison2013}
{Harrison}, F.~A., {Craig}, W.~W., {Christensen}, F.~E., {et~al.} 2013, \bibinfo{title}{{The Nuclear Spectroscopic Telescope Array (NuSTAR) High-energy X-Ray Mission},} \apj, 770, 103, \dodoi{10.1088/0004-637X/770/2/103}

\bibitem[{G. {Hasinger} \& M. {van der Klis}(1989){Hasinger} \& {van der Klis}}]{Hasinger1989}
{Hasinger}, G., \& {van der Klis}, M. 1989, \bibinfo{title}{{Two patterns of correlated X-ray timing and spectral behaviour in low-mass X-ray binaries.},} \aap, 225, 79

\bibitem[{J.~D. Hunter(2007)Hunter}]{Hunter2007}
Hunter, J.~D. 2007, \bibinfo{title}{Matplotlib: A 2D graphics environment,} Computing in Science \& Engineering, 9, 90, \dodoi{10.1109/MCSE.2007.55}

\bibitem[{R. {Iaria} {et~al.}(2009){Iaria}, {D'A{\'\i}}, {di Salvo}, {Robba}, {Riggio}, {Papitto}, \& {Burderi}}]{Iaria2009}
{Iaria}, R., {D'A{\'\i}}, A., {di Salvo}, T., {et~al.} 2009, \bibinfo{title}{{A ionized reflecting skin above the accretion disk of GX 349+2},} \aap, 505, 1143, \dodoi{10.1051/0004-6361/200911936}

\bibitem[{U. {Kashyap} {et~al.}(2023){Kashyap}, {Chakraborty}, {Bhattacharyya}, \& {Ram}}]{Kashyap2023}
{Kashyap}, U., {Chakraborty}, M., {Bhattacharyya}, S., \& {Ram}, B. 2023, \bibinfo{title}{{Broad-band spectro-temporal investigation of neutron star low-mass X-ray binary GX 349+2},} \mnras, 523, 2788, \dodoi{10.1093/mnras/stad1606}

\bibitem[{E. {Kuulkers} \& M. {van der Klis}(1998){Kuulkers} \& {van der Klis}}]{Kuulkers1998}
{Kuulkers}, E., \& {van der Klis}, M. 1998, \bibinfo{title}{{GX349+2 (ScoX-2): an odd-ball among the Z sources},} \aap, 332, 845, \dodoi{10.48550/arXiv.astro-ph/9712311}

\bibitem[{F. {La Monaca} {et~al.}(2024{\natexlab{a}}){La Monaca}, {Di Marco}, {Poutanen}, {Bachetti}, {Motta}, {Papitto}, {Pilia}, {Xie}, {Bianchi}, {Bobrikova}, {Costa}, {Deng}, {Ge}, {Illiano}, {Jia}, {Krawczynski}, {Lai}, {Liu}, {Mastroserio}, {Muleri}, {Rankin}, {Soffitta}, {Veledina}, {Ambrosino}, {Del Santo}, {Chen}, {Garcia}, {Kaaret}, {Russell}, {Wei}, {Zhang}, {Zuo}, {Arzoumanian}, {Cocchi}, {Gnarini}, {Farinelli}, {Gendreau}, {Ursini}, {Weisskopf}, {Zane}, {Agudo}, {Antonelli}, {Baldini}, {Baumgartner}, {Bellazzini}, {Bongiorno}, {Bonino}, {Brez}, {Bucciantini}, {Capitanio}, {Castellano}, {Cavazzuti}, {Chen}, {Ciprini}, {De Rosa}, {Del Monte}, {Di Gesu}, {Di Lalla}, {Donnarumma}, {Doroshenko}, {Dov{\v{c}}iak}, {Ehlert}, {Enoto}, {Evangelista}, {Fabiani}, {Ferrazzoli}, {Gunji}, {Hayashida}, {Heyl}, {Iwakiri}, {Jorstad}, {Karas}, {Kislat}, {Kitaguchi}, {Kolodziejczak}, {Latronico}, {Liodakis}, {Maldera}, {Manfreda}, {Marin}, {Marinucci}, {Marscher}, {Marshall}, {Massaro}, {Matt}, {Mitsuishi},
  {Mizuno}, {Negro}, {Ng}, {O'Dell}, {Omodei}, {Oppedisano}, {Pavlov}, {Peirson}, {Perri}, {Pesce-Rollins}, {Petrucci}, {Possenti}, {Puccetti}, {Ramsey}, {Ratheesh}, {Roberts}, {Romani}, {Sgr{\`o}}, {Slane}, {Spandre}, {Swartz}, {Tamagawa}, {Tavecchio}, {Taverna}, {Tawara}, {Tennant}, {Thomas}, {Tombesi}, {Trois}, {Tsygankov}, {Turolla}, {Vink}, {Wu}, \& {IXPE Collaboration}}]{LaMonaca2024}
{La Monaca}, F., {Di Marco}, A., {Poutanen}, J., {et~al.} 2024{\natexlab{a}}, \bibinfo{title}{{Highly Significant Detection of X-Ray Polarization from the Brightest Accreting Neutron Star Sco X-1},} \apjl, 960, L11, \dodoi{10.3847/2041-8213/ad132d}

\bibitem[{F. {La Monaca} {et~al.}(2024{\natexlab{b}}){La Monaca}, {Di Marco}, {Poutanen}, {Bachetti}, {Motta}, {Papitto}, {Pilia}, {Xie}, {Bianchi}, {Bobrikova}, {Costa}, {Deng}, {Ge}, {Illiano}, {Jia}, {Krawczynski}, {Lai}, {Liu}, {Mastroserio}, {Muleri}, {Rankin}, {Soffitta}, {Veledina}, {Ambrosino}, {Del Santo}, {Chen}, {Garcia}, {Kaaret}, {Russell}, {Wei}, {Zhang}, {Zuo}, {Arzoumanian}, {Cocchi}, {Gnarini}, {Farinelli}, {Gendreau}, {Ursini}, {Weisskopf}, {Zane}, {Agudo}, {Antonelli}, {Baldini}, {Baumgartner}, {Bellazzini}, {Bongiorno}, {Bonino}, {Brez}, {Bucciantini}, {Capitanio}, {Castellano}, {Cavazzuti}, {Chen}, {Ciprini}, {De Rosa}, {Del Monte}, {Di Gesu}, {Di Lalla}, {Donnarumma}, {Doroshenko}, {Dov{\v{c}}iak}, {Ehlert}, {Enoto}, {Evangelista}, {Fabiani}, {Ferrazzoli}, {Gunji}, {Hayashida}, {Heyl}, {Iwakiri}, {Jorstad}, {Karas}, {Kislat}, {Kitaguchi}, {Kolodziejczak}, {Latronico}, {Liodakis}, {Maldera}, {Manfreda}, {Marin}, {Marinucci}, {Marscher}, {Marshall}, {Massaro}, {Matt}, {Mitsuishi},
  {Mizuno}, {Negro}, {Ng}, {O'Dell}, {Omodei}, {Oppedisano}, {Pavlov}, {Peirson}, {Perri}, {Pesce-Rollins}, {Petrucci}, {Possenti}, {Puccetti}, {Ramsey}, {Ratheesh}, {Roberts}, {Romani}, {Sgr{\`o}}, {Slane}, {Spandre}, {Swartz}, {Tamagawa}, {Tavecchio}, {Taverna}, {Tawara}, {Tennant}, {Thomas}, {Tombesi}, {Trois}, {Tsygankov}, {Turolla}, {Vink}, {Wu}, \& {IXPE Collaboration}}]{LaMonaca2024x}
{La Monaca}, F., {Di Marco}, A., {Poutanen}, J., {et~al.} 2024{\natexlab{b}}, \bibinfo{title}{{Highly Significant Detection of X-Ray Polarization from the Brightest Accreting Neutron Star Sco X-1},} \apjl, 960, L11, \dodoi{10.3847/2041-8213/ad132d}

\bibitem[{L.-X. {Li} {et~al.}(2009){Li}, {Narayan}, \& {McClintock}}]{Li2009}
{Li}, L.-X., {Narayan}, R., \& {McClintock}, J.~E. 2009, \bibinfo{title}{{Inferring the Inclination of a Black Hole Accretion Disk from Observations of its Polarized Continuum Radiation},} \apj, 691, 847, \dodoi{10.1088/0004-637X/691/1/847}

\bibitem[{V. {Loktev} {et~al.}(2022){Loktev}, {Veledina}, \& {Poutanen}}]{Loktev2022}
{Loktev}, V., {Veledina}, A., \& {Poutanen}, J. 2022, \bibinfo{title}{{Analytical techniques for polarimetric imaging of accretion flows in the Schwarzschild metric},} \aap, 660, A25, \dodoi{10.1051/0004-6361/202142360}

\bibitem[{K. {Mitsuda} {et~al.}(1989){Mitsuda}, {Inoue}, {Nakamura}, \& {Tanaka}}]{Mitsuda1989}
{Mitsuda}, K., {Inoue}, H., {Nakamura}, N., \& {Tanaka}, Y. 1989, \bibinfo{title}{{Luminosity-related changes of the energy spectrum of X 1608-522.},} \pasj, 41, 97

\bibitem[{K. {Mitsuda} {et~al.}(1984){Mitsuda}, {Inoue}, {Koyama}, {Makishima}, {Matsuoka}, {Ogawara}, {Shibazaki}, {Suzuki}, {Tanaka}, \& {Hirano}}]{Mitsuda1984}
{Mitsuda}, K., {Inoue}, H., {Koyama}, K., {et~al.} 1984, \bibinfo{title}{{Energy spectra of low-mass binary X-ray sources observed from Tenma.},} \pasj, 36, 741

\bibitem[{J. {Nishimura} {et~al.}(1986){Nishimura}, {Mitsuda}, \& {Itoh}}]{Nishimura1986}
{Nishimura}, J., {Mitsuda}, K., \& {Itoh}, M. 1986, \bibinfo{title}{{Comptonization of soft X-ray photons in an optically thin hot plasma},} \pasj, 38, 819

\bibitem[{J. {Rankin} {et~al.}(2024){Rankin}, {La Monaca}, {Di Marco}, {Poutanen}, {Bobrikova}, {Kravtsov}, {Muleri}, {Pilia}, {Veledina}, {Fender}, {Kaaret}, {Kim}, {Marinucci}, {Marshall}, {Papitto}, {Tennant}, {Tsygankov}, {Weisskopf}, {Wu}, {Zane}, {Ambrosino}, {Farinelli}, {Gnarini}, {Agudo}, {Antonelli}, {Bachetti}, {Baldini}, {Baumgartner}, {Bellazzini}, {Bianchi}, {Bongiorno}, {Bonino}, {Brez}, {Bucciantini}, {Capitanio}, {Castellano}, {Cavazzuti}, {Chen}, {Ciprini}, {Costa}, {De Rosa}, {Del Monte}, {Di Gesu}, {Di Lalla}, {Donnarumma}, {Doroshenko}, {Dov{\v{c}}iak}, {Ehlert}, {Enoto}, {Evangelista}, {Fabiani}, {Ferrazzoli}, {Garcia}, {Gunji}, {Hayashida}, {Heyl}, {Iwakiri}, {Jorstad}, {Karas}, {Kislat}, {Kitaguchi}, {Kolodziejczak}, {Krawczynski}, {Latronico}, {Liodakis}, {Maldera}, {Manfreda}, {Marin}, {Marscher}, {Massaro}, {Matt}, {Mitsuishi}, {Mizuno}, {Negro}, {Ng}, {O'Dell}, {Omodei}, {Oppedisano}, {Pavlov}, {Peirson}, {Perri}, {Pesce-Rollins}, {Petrucci}, {Possenti}, {Puccetti}, {Ramsey},
  {Ratheesh}, {Roberts}, {Romani}, {Sgr{\`o}}, {Slane}, {Soffitta}, {Spandre}, {Swartz}, {Tamagawa}, {Tavecchio}, {Taverna}, {Tawara}, {Thomas}, {Tombesi}, {Trois}, {Turolla}, {Vink}, \& {Xie}}]{Rankin2024}
{Rankin}, J., {La Monaca}, F., {Di Marco}, A., {et~al.} 2024, \bibinfo{title}{{X-Ray Polarized View of the Accretion Geometry in the X-Ray Binary Circinus X-1},} \apjl, 961, L8, \dodoi{10.3847/2041-8213/ad1832}

\bibitem[{P. {Romano} {et~al.}(2006){Romano}, {Campana}, {Chincarini}, {Cummings}, {Cusumano}, {Holland}, {Mangano}, {Mineo}, {Page}, {Pal'Shin}, {Rol}, {Sakamoto}, {Zhang}, {Aptekar}, {Barbier}, {Barthelmy}, {Beardmore}, {Boyd}, {Burrows}, {Capalbi}, {Fenimore}, {Frederiks}, {Gehrels}, {Giommi}, {Goad}, {Godet}, {Golenetskii}, {Guetta}, {Kennea}, {La Parola}, {Malesani}, {Marshall}, {Moretti}, {Nousek}, {O'Brien}, {Osborne}, {Perri}, \& {Tagliaferri}}]{Romano2006}
{Romano}, P., {Campana}, S., {Chincarini}, G., {et~al.} 2006, \bibinfo{title}{{Panchromatic study of GRB 060124: from precursor to afterglow},} \aap, 456, 917, \dodoi{10.1051/0004-6361:20065071}

\bibitem[{L. {Titarchuk}(1994){Titarchuk}}]{Titarchuk1994}
{Titarchuk}, L. 1994, \bibinfo{title}{{Generalized Comptonization Models and Application to the Recent High-Energy Observations},} \apj, 434, 570, \dodoi{10.1086/174760}

\bibitem[{F. {Ursini} {et~al.}(2024){Ursini}, {Gnarini}, {Capitanio}, {Bobrikova}, {Cocchi}, {Di Marco}, {Fabiani}, {Farinelli}, {La Monaca}, {Rankin}, {Saade}, \& {Poutanen}}]{Ursini2024}
{Ursini}, F., {Gnarini}, A., {Capitanio}, F., {et~al.} 2024, \bibinfo{title}{{The IXPE View of Neutron Star Low-Mass X-ray Binaries},} Galaxies, 12, 43, \dodoi{10.3390/galaxies12040043}

\bibitem[{J. {van Paradijs} \& J.~E. {McClintock}(1994){van Paradijs} \& {McClintock}}]{vanParadijs1994}
{van Paradijs}, J., \& {McClintock}, J.~E. 1994, \bibinfo{title}{{Absolute visual magnitudes of low-mass X-ray binaries.},} \aap, 290, 133

\bibitem[{M.~C. {Weisskopf} {et~al.}(2016){Weisskopf}, {Ramsey}, {O'Dell}, {Tennant}, {Elsner}, {Soffitta}, {Bellazzini}, {Costa}, {Kolodziejczak}, {Kaspi}, {Muleri}, {Marshall}, {Matt}, \& {Romani}}]{Weisskopf2016}
{Weisskopf}, M.~C., {Ramsey}, B., {O'Dell}, S., {et~al.} 2016, in Society of Photo-Optical Instrumentation Engineers (SPIE) Conference Series, Vol. 9905, Space Telescopes and Instrumentation 2016: Ultraviolet to Gamma Ray, ed. J.-W.~A. {den Herder}, T.~{Takahashi}, \& M.~{Bautz}, 990517, \dodoi{10.1117/12.2235240}

\bibitem[{M.~C. {Weisskopf} {et~al.}(2022){Weisskopf}, {Soffitta}, {Baldini}, {Ramsey}, {O'Dell}, {Romani}, {Matt}, {Deininger}, {Baumgartner}, {Bellazzini}, {Costa}, {Kolodziejczak}, {Latronico}, {Marshall}, {Muleri}, {Bongiorno}, {Tennant}, {Bucciantini}, {Dovciak}, {Marin}, {Marscher}, {Poutanen}, {Slane}, {Turolla}, {Kalinowski}, {Di Marco}, {Fabiani}, {Minuti}, {La Monaca}, {Pinchera}, {Rankin}, {Sgro'}, {Trois}, {Xie}, {Alexander}, {Allen}, {Amici}, {Andersen}, {Antonelli}, {Antoniak}, {Attin{\`a}}, {Barbanera}, {Bachetti}, {Baggett}, {Bladt}, {Brez}, {Bonino}, {Boree}, {Borotto}, {Breeding}, {Brienza}, {Bygott}, {Caporale}, {Cardelli}, {Carpentiero}, {Castellano}, {Castronuovo}, {Cavalli}, {Cavazzuti}, {Ceccanti}, {Centrone}, {Citraro}, {D'Amico}, {D'Alba}, {Di Gesu}, {Del Monte}, {Dietz}, {Di Lalla}, {Persio}, {Dolan}, {Donnarumma}, {Evangelista}, {Ferrant}, {Ferrazzoli}, {Ferrie}, {Footdale}, {Forsyth}, {Foster}, {Garelick}, {Gunji}, {Gurnee}, {Head}, {Hibbard}, {Johnson}, {Kelly}, {Kilaru},
  {Lefevre}, {Roy}, {Loffredo}, {Lorenzi}, {Lucchesi}, {Maddox}, {Magazzu}, {Maldera}, {Manfreda}, {Mangraviti}, {Marengo}, {Marrocchesi}, {Massaro}, {Mauger}, {McCracken}, {McEachen}, {Mize}, {Mereu}, {Mitchell}, {Mitsuishi}, {Morbidini}, {Mosti}, {Nasimi}, {Negri}, {Negro}, {Nguyen}, {Nitschke}, {Nuti}, {Onizuka}, {Oppedisano}, {Orsini}, {Osborne}, {Pacheco}, {Paggi}, {Painter}, {Pavelitz}, {Pentz}, {Piazzolla}, {Perri}, {Pesce-Rollins}, {Peterson}, {Pilia}, {Profeti}, {Puccetti}, {Ranganathan}, {Ratheesh}, {Reedy}, {Root}, {Rubini}, {Ruswick}, {Sanchez}, {Sarra}, {Santoli}, {Scalise}, {Sciortino}, {Schroeder}, {Seek}, {Sosdian}, {Spandre}, {Speegle}, {Tamagawa}, {Tardiola}, {Tobia}, {Thomas}, {Valerie}, {Vimercati}, {Walden}, {Weddendorf}, {Wedmore}, {Welch}, {Zanetti}, \& {Zanetti}}]{Weisskopf2022}
{Weisskopf}, M.~C., {Soffitta}, P., {Baldini}, L., {et~al.} 2022, \bibinfo{title}{{The Imaging X-Ray Polarimetry Explorer (IXPE): Pre-Launch},} Journal of Astronomical Telescopes, Instruments, and Systems, 8, 026002, \dodoi{10.1117/1.JATIS.8.2.026002}

\bibitem[{N.~E. {White} {et~al.}(1988){White}, {Stella}, \& {Parmar}}]{White1988}
{White}, N.~E., {Stella}, L., \& {Parmar}, A.~N. 1988, \bibinfo{title}{{The X-Ray Spectral Properties of Accretion Disks in X-Ray Binaries},} \apj, 324, 363, \dodoi{10.1086/165901}

\bibitem[{J. {Wilms} {et~al.}(2000){Wilms}, {Allen}, \& {McCray}}]{Wilms2000}
{Wilms}, J., {Allen}, A., \& {McCray}, R. 2000, \bibinfo{title}{{On the Absorption of X-Rays in the Interstellar Medium},} \apj, 542, 914, \dodoi{10.1086/317016}

\end{thebibliography}
\bibliographystyle{aasjournalv7}



\end{document}